\documentclass[showpacs, twocolumn]{revtex4-1}
\usepackage{graphicx}

\begin{document}

\title{Two-dimensional dipolar Bose gas with the roton-maxon excitation spectrum}

\author{Abdel\^{a}ali Boudjema\^{a}$^{1,2}$ and G.V. Shlyapnikov$^{2,3}$}

\affiliation{$^1$Department of Physics, Faculty of Sciences, Hassiba Benbouali University of Chlef P.O. Box 151, 02000, Chlef, Algeria
\\$^2$Laboratoire de Physique Th\'{e}orique et Mod\`{e}les Statistiques, CNRS and Universit\'{e} Paris Sud, UMR8626, 91405 Orsay, France
\\ $^3$Van der Waals-Zeeman Institute, University of Amsterdam, Science Park 904, 1098 XH Amsterdam, The Netherlands}

\begin{abstract}
We discuss fluctuations in a dilute two-dimensional Bose-condensed dipolar gas, which has a roton-maxon character of the excitation spectrum.
We calculate the density-density correlation function, fluctuation corrections to the chemical potential, compressibility, and the normal (superfluid) fraction. It is shown that the presence of the roton strongly enhances fluctuations of the density, and we establish the validity criterion of the Bogoliubov approach. At $T=0$ the condensate depletion becomes significant if
the roton minimum is sufficiently close to zero. At finite temperatures exceeding the roton energy, the effect of thermal fluctuations is stronger and it may lead to a large normal fraction of the gas and compressibility. 
 
\end{abstract}


\maketitle

%\section{Introduction}
%%%%%%%%%%%%

In the last decade, ultracold gases of dipolar particles, which include atoms with a large magnetic moment and polar molecules,
attracted a great deal of interest \cite{Baranov,Pfau,Carr,Pupillo2012}. Being electrically or magnetically polarized such particles 
interact with each other via long-range anisotropic dipole-dipole forces, which drastically changes the nature of quantum degenerate
regimes. Experiments with chromium atoms (magnetic moment $6\mu_B$) \cite{Pfau}, together with theoretical studies \cite{Santos,Eberlein}, have 
revealed the dependence of the shape and stability diagram of trapped dipolar Bose-Einstein condensates on the trapping geometry and interaction
strength. They initiated spinor physics with quantum dipoles \cite{Santos2,pasqu}, and now also dysprosium \cite{ming} and erbium \cite{erbium} atoms (magnetic moments $10\mu_B$ and $7\mu_B$, respectively) 
have entered the game.
Recently, fascinating prospects for the observation of novel quantum phases have been opened by the creation of ultracold clouds 
of polar molecules and cooling them to almost quantum degeneracy \cite{kk,Carr}. In this case, the dipole-dipole forces can
be orders of magnitude larger, and one has a possibility to manipulate the molecules making use of their rotational degrees of 
freedom.

For dipolar Bose-condensed gases, one of the key issues was related  to the presence of the roton-maxon character of the excitation spectrum and to the possibility of obtaining supersolid states in which the condensate wavefunction is a
superposition of a uniform background and a lattice structure. The roton-maxon structure of the spectrum has been first predicted for strongly
pancaked Bose-Einstein condensates \cite{gora}. However, the idea to obtain a supersolid state when the roton touches zero and uniform BEC
becomes unstable, did not succeed because of the collapse of the system \cite{gora2,cooper}. Since that time, several proposals have been
made for the creation of supersolid states with bosons \cite{prok}. They rely on the potential of interatomic interaction which is flat at short distances and decays at large separations, and the results then indicate the presence of dense supersolid clusters \cite{cint}.

This activity brought in analogies with liquid helium, where the 
studies of the roton-maxon spectrum and the attempts to observe experimentally the supersolid state have spanned for decades \cite{bali,Noz}
after the early theoretical prediction \cite{andr}. However, the most credible claim for the observation of the supersolid \cite{kim} is now withdrawn 
\cite{kim1}. It is worth mentioning that the old idea of obtaining a stable density-modulated state (supersolid) in superfluid helium moving with a supercritical velocity \cite{Pit84} is now being discussed in a general context of superfluidity in Bose gases flowing with velocities larger than the Landau critical velocity \cite{Baym}.

The roton-maxon character of the excitation spectrum was also attracting large attention by itself. In relation to
liquid helium, it has been discussed how the position of the roton minimum influences the phenomenon of superfluidity \cite{Noz}.
In the context of dipolar bosons in two dimensions, numerical calculations of zero temperature phase diagram \cite{mora,bush,Astr} found that the reduction
of the condensed fraction with an increase in the density can be attributed to the appearance of the roton minimum. Finite temperature
Monte Carlo calculations \cite{prok3} have revealed that the rotonization of the spectrum can decrease the Kosterlitz-Thouless superfluid
transition temperature. Although the calculations \cite{mora,bush,Astr,prok3} were focused on fairly high densities, they raised the question of applicability of the Bogoliubov approach for dipolar bosons \cite{Astr}. At the same time, there was 
an extended activity on the static and dynamical properties of dilute trapped dipolar Bose-Einstein condensates on the basis of the Bogoliubov approach \cite{uwe,bohn1,bohn2,bohn3}. 

It is therefore instructive to identify the validity criterion of the Bogoliubov approach for Bose-condensed dipolar gases with the roton-maxon excitation spectrum, 
and this is the subject of the present paper. We show that at zero temperature the
density fluctuations originating from the presence of the roton minimum, lead to a significant depletion of the condensate
and modify thermodynamic quantities if the roton minimum is sufficiently close to zero. 
At finite temperatures exceeding the roton energy, thermal density fluctuations may have a much stronger influence, leading to a large increase of the normal fraction and compressibility.

We consider a dilute Bose-condensed gas of dipolar bosons (tightly) confined in one direction $(z)$ to zero point oscillations and assume that in the 
$x,y$ plane the translational motion is free (see Fig.1). The dipole moments are oriented perpendicularly to the $x,y$ plane, which for electric dipoles (polar molecules) can be done applying an electric field, and for magnetic atoms by using a magnetic field. In this quasi-2D geometry, at large interparticle separations $r$ the interaction potential is  
\begin{equation}\label{dd}
V(r) = \frac{d^2}{r^3}=\frac{\hbar^2r_*}{mr^3},
\end{equation}
with $d$ being the dipole moment, $m$ the particle mass, and $r_*=md^2/\hbar^2$ the characteristic dipole-dipole distance. The short-range part of the potential is assumed to be such that there is a roton-maxon excitation spectrum. 
\begin{figure}[htb1]
\includegraphics[scale=0.5]{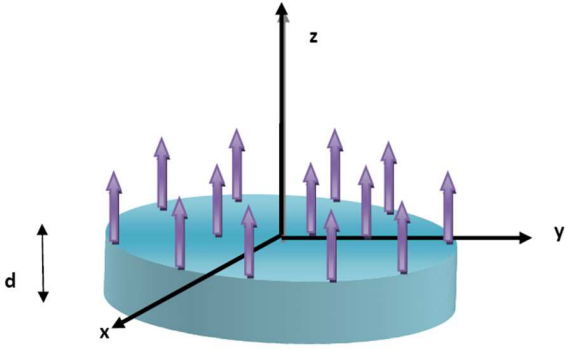}
\caption{Dipolar Bose-Einstein condensate tightly confined in one direction.}
\end{figure}
In the ultracold limit where the particle momenta satisfy the inequality $kr_*\ll1$, the off-shell scattering amplitude defined as
$f({\vec k},{\vec k'})=\int \exp(-i{\vec k'} {\vec r'}) V(r)\psi_{\vec k}({\vec r})d^2r$ ($\psi_{\vec k}({\vec r})$ is the wavefunction 
of the relative motion with momentum $\vec{k}$), is given by (see \cite{Pikgora} and refs. therein):
\begin{equation}\label{scam1}
 f({\vec k},{\vec k'})=\frac{\hbar^2}{m}\left[\frac{2\pi}{\ln(\kappa/k)+i\pi/2}-2\pi r_*\vert \vec k-\vec k'\vert\right], 
\end{equation}
where we omit higher order terms in $k$. 
The second term in the right hand side represents the so-called anomalous contribution coming from distances of the order of the de Broglie wavelength of particles \cite{Landlif}. It takes into account all partial waves and is obtained using a perturbative approach in $V(r)$. 
The first term in the right hand side of Eq.(\ref{scam1}) describes the short-range contribution. It is obtained by putting $k'=0$
and proceeding along the lines of the 2D scattering theory \cite{Landlif}. The parameter $\kappa$ depends on the behavior of $V(r)$ at short distances.
In the quasi-2D geometry it also depends on the confinement length in the $z$-direction, $l_0=\sqrt{\hbar/m\omega_0}$,
where $\omega_0$ is the confinement frequency.
One then can express $\kappa$ through the 3D coupling constant $g_{3D}$.
If the 3D s-wave scattering length, $ a_{3D}=mg_{3D}/4\pi \hbar^2\ll l_0$, then $\kappa$ is exponentially small 
\cite{ petr1,petr2} and we may omit the k-dependence under logarithm in Eq.(\ref{scam1}), as well as $i\pi/2$ in the denominator of the first term. This gives $f({\vec k},{\vec k'})=g(1-C\vert \vec k-\vec k'\vert)$, where the 2D short-range coupling constant is $g=g_{3D}/\sqrt{2}l_0$ and $C =2\pi \hbar^2r_*/mg=2\pi d^2/g$.
Employing this result in the secondly quantized Hamiltonian \cite{gora2,cooper}, we obtain
\begin{eqnarray}\label{he3}
\!\!\!\!\hat H\!\!=\!\!\sum_{\vec k}\!E_k\hat a^\dagger_{\vec k}\hat a_{\vec k}\!+\!\frac{g}{2S}\!\!\sum_{\vec k,\vec q,\vec p}\!\!
(1\!\!-\!C\vert \vec q\!-\!\vec p\vert)\hat a^\dagger_{\vec k\!+\!\vec q} \hat a^\dagger_{\vec k\!-\!\vec q}\hat a_{\vec k\!+\!\vec p}\hat a_{\vec k\!-\!\vec p}
\end{eqnarray}
where $S$ is the surface area, $E_k=\hbar^2k^2/2m$, and $\hat a_{\bf k}^\dagger$, $\hat a_{\bf k}$ are the creation and annihilation operators of particles.
At zero temperature there is a true Bose-Einstein condensate in 2D, and we may use the standard Bogoliubov approach.
Assuming the weakly interacting regime where $mg/2\pi\hbar^2\ll 1$ and $r_*\ll \xi$, with $\xi=\hbar/\sqrt{mng}$ being the healing length,
we reduce the Hamiltonian (\ref{he3}) to a bilinear form, use the Bogoliubov transformation $\hat a^\dagger_{\vec k}= u_k \hat b^\dagger_{\vec k}-v_k \hat b_{-\vec k}$,  
and obtain the diagonal form $\hat H = E_0+\sum_{\vec k} \varepsilon_k\hat b^\dagger_{\vec k}\hat b_{\vec k}$ in terms of operators
$b^\dagger_{\vec k}$, $\hat b_{\vec k}$ of elementary excitations. The Bogoliubov functions $ u_k,v_k$ are expressed in a standard way:
$ u_k,v_k=(\sqrt{\varepsilon_k/E_k}\pm\sqrt{E_k/\varepsilon_k})/2$, and  the Bogoluibov excitation energy is given by 
$\varepsilon_k=\sqrt{E_k^{2}+2ngE_k(1-Ck)}$.

To zero order the chemical potential is $\mu=ng$.
For small momenta the excitations are sound waves, $\varepsilon_k=\sqrt{ng/m}k$. The dependence of $\varepsilon_k$
on $k$ remains monotonic with increasing $k$ if $C\leq \sqrt{8}\xi/3$ (see Fig.2 ).
For the constant $C$ in the interval 
\begin{equation}\label{pump3} 
\frac{\sqrt{8}}{3}\xi\leq C\leq\xi,
\end{equation}
the excitation spectrum has a roton-maxon structure.
It is then convenient to represent $\varepsilon_k$ in the form:
\begin{equation}\label{pump4} 
\varepsilon_k= \frac{\hbar^2 k}{2m}\sqrt{ (k-k_r)^2 +k_{\Delta}^2},
\end{equation}
where $k_r=2C/\xi^2$ and $k_{\Delta}=\sqrt{4/\xi^2-k_r^2}$.
If the roton is close to zero, then $k_r$ is the position of the roton, 
and 
\begin{equation}    \label{Delta}
\!\!\Delta\!=\!\hbar^2 k_rk_{\Delta}/2m\!=\!2ngC\sqrt{mng/\hbar^2-C^2(mng/\hbar^2)^2}\!\!
\end{equation} 
is the height of the roton minimum (see Fig.2).
For $C=\xi$ the roton minimum touches zero, and at larger $C$ the uniform Bose condensate becomes dynamically unstable.
%This is reflected in the appearance of imaginary excitation energies $\varepsilon_k$.

It should be noted that the coupling constant $g$ can be tuned by using Feshbach resonances or by modifying the frequency of the 
tight confinement $\omega_0$. Therefore, although the range of $C$ given by Eq.(\ref{pump3}) is rather narrow, it can be reached without serious
difficulties. The condition $C=2\pi d^2/g=\xi$ is reduced to $(mg/2\pi\hbar^2)=a_{3D}/\sqrt{2}l_0\simeq 2\pi nr_*^2$. For dysprosium atoms we have the dipole-dipole distance $r_*\simeq 200$ \AA, and at 2D densities $\sim 10^9$ cm$^{-2}$ the roton-maxon spectrum is realized for the 3D scattering length $a_{3D}$ of several tens of angstroms at the frequency of the tight confinement of $10$ kHz leading to the confinement length $l_0$ about 1000 \AA. 
 
%\begin{widetext}
%\begin{equation}\label{he2}
%\hat H = \int d^2r \, \hat \psi^\dagger(\mathbf{r}) \left\{\frac{-\hbar^2 }{2m}\Delta_\mathbf{r}+V(\mathbf{r})+
%\frac{1}{2}\int d^2r^\prime\, \hat\psi^\dagger (\mathbf{r^\prime}) V(\mathbf{r}-\mathbf{r^\prime})\hat\psi^\dagger(\mathbf{r^\prime})\right\}\hat\psi(\mathbf{r}) ,
%\end{equation}
%\end{widetext}
%where $\psi^\dagger$ and $\psi$ denote respectevly the usual creation and annihilation field operators, 
%$V(\mathbf{r}-\mathbf{r^\prime})=g\delta(\mathbf{r}-\mathbf{r^\prime})+ d^2 \frac{1-3\cos^2\theta}{\vert r-r^\prime\vert^3}\ $ is the interaction potential
%with $d^2$ is the dipole moment and $\theta$ being the angel between the vector $\vert r-r^\prime\vert$ and the direction of dipoles $(z)$.

%we use the Bogoluibov transformation and separtaing out the condensate from the field operator 
%$\hat\psi=\hat\psi^\prime+\psi_0 $, where $\hat\psi^\prime= \sum_{\vec k} u_k\hat a_k\exp(\frac{-i\varepsilon_k t}{\hbar})-v_k\hat a^\dagger_k \exp(\frac{i\varepsilon_k t}{\hbar})$ 
%accounts for the non-condensed densty.
\begin{figure}[htb]
\includegraphics[scale=0.7, angle=0]{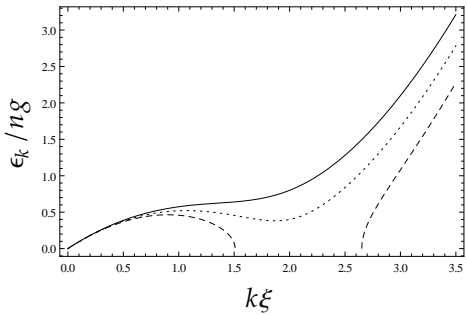}
\caption{Excitation energy $\varepsilon_k$ of the quasi-2D dipolar BEC as a function of momentum $k$ for several values of $k_r$. The solid curve ($k_r\xi=1.84$) shows a monotonic dependence $\varepsilon_{k}$, 
the dotted curve ($k_r\xi=1.96$) is $\varepsilon_k$ with the roton-maxon structure, and the dashed curve ($k_r\xi=2.08$) corresponds to dynamically unstable BEC.}
\end{figure}

%\section{Density fluctuations and condensate depletion at $T=0$}

The Bogoliubov approach assumes that the density and phase fluctuations are small. In the 2D case at $T=0$, the presence of the roton does not significantly change the phase fluctuations and they remain small. However, the situation with the density fluctuations is different. Writing the operator of the density fluctuations as (see \cite{LL9}) $\delta\hat n=\sqrt{n}\sum_{\bf k}(u_k-v_k)\exp(i{\bf kr})\hat b_{\bf k}+h.c.$, 
we obtain for the density-density correlation function:
\begin{equation}     \label{deltan1}
\frac{\langle \delta\hat n({\bf r})\delta\hat (0)\rangle}{n^2}=\frac{1}{n}\int\frac{d^2k}{(2\pi)^2}\frac{E_k}{\varepsilon_k}(1+2N_k)\exp(i{\bf kr}),
\end{equation}
where $N_k=[\exp(\varepsilon_k/T)-1]^{-1}$ are occupation numbers for the excitations. It is instructive to single out the roton contribution to the correlation function (\ref{deltan1}). Asuming that the roton is close to zero and the roton energy is $\Delta\ll ng$, we have the cofficient $C$  close to $\xi$, and $k_r\simeq 2/\xi$. Then, using Eqs.(\ref{pump4}) and (\ref{deltan1}), for the contribution of momenta near the roton minimum at $T=0$ we obtain:
%\begin{equation}     \label{deltan2}
%\frac{\langle \delta\hat n({\bf r})\delta\hat (0)\rangle_r}{n^2}=\frac{k_r^2}{2\pi n}\ln\left(\frac{k_r}{k_{\Delta}}\right)J_0(k_rr);\,\,\,\,\,k_%{\Delta}r\ll 1,
%\end{equation}
\begin{equation}       \label{deltanfin}
\!\!\frac{\langle \delta\hat n({\bf r})\delta\hat n(0)\rangle_r}{n^2}\!=\!\frac{2mg}{\pi\hbar^2}\ln\left(\frac{2ng}{\Delta}\right)J_0(2r/\xi);\,\,\Delta\ll ng,
\end{equation}
where $J_0$ is the Bessel function.

We thus see that the density fluctuations grow logarithmically when the roton minimum is approaching zero and they can become strong for very small $\Delta$.
In this case they lead to a significant depletion of the condensate. The non-condensed density of particles is 
$n'=\int v_k^2 d^2k/(2\pi)^2$ and the integral over $dk$ is logarithmically divergent at large momenta because of the dipolar contribution to the interaction strength, $-gCk$. However, this form of the dipole-dipole contribution is valid only for $k\ll 1/r_*$. We thus may put a high momentum cut-off $1/r_*$, which leads to (see Fig.3):
\begin{equation}     \label{nonBEC1}
\!\!\!\!\frac{n'}{n}\!\!=\!\!\frac{mg}{4\pi\hbar^2}\!\left[1\!\!-\!k_r\xi\!-\!\frac{3(k_r\xi)^2}{4}\!+\!\frac{(k_r\xi)^2}{2}\ln\!\!\left(\!\frac{\xi}{r_*(2\!\!-\!k_r\xi)}\!\!\right)\!\right]\!.\!\!\!\!
\end{equation}
In the absence of the dipole-dipole interaction ($r_*=0$ and $k_r=0$) we recover the usual result for the 2D BEC with short-range 
interparticle repulsion, $n'=n(mg/4\pi\hbar^2)$. For $\Delta\ll ng$ we have $(2-k_r\xi)\simeq (k_{\Delta}\xi)^2/4$ and Eq.(\ref{nonBEC1})
transforms to
\begin{equation}    \label{nonBECfin}
\frac{n'}{n}\simeq \frac{mg}{\pi\hbar^2}\ln\left(\frac{2ng}{\Delta}\zeta\right);\,\,\,\,\,\,\Delta\ll ng,
\end{equation}
where $\zeta=\sqrt{2\pi\hbar^2/e^2 mg}$.

As we see from Eqs.~(\ref{deltanfin}) and (\ref{nonBECfin}), for the roton minimum close to zero a small condensate depletion and small fluctuations of the density require the inequality $(mg/\pi\hbar^2)\ln(2ng/\Delta)\ll 1$. It differs only by a logarithmic factor $\ln(2ng/\Delta)$ from the small 
parameter of the theory, $(mg/2\pi\hbar^2)\ll 1$, in the absence of the roton. 
%Thus, the inequality (\ref{ineq1}) indicates that at zero temperature only for $\Delta$ extremely close to zero the fluctuations are so strong that the Bogoliubov approach fails (see Fig.3).
The same logarithmic factor appears in the fluctuation correction to the chemical potential and in the one-body density matrix $g_1(r)=\langle\hat\Psi^{\dagger}({\bf r})\hat\Psi(0)\rangle$,
where $\hat\Psi({\bf r})$ is the field operator. Assuming that the roton minimum is close to zero and taking into account only the contribution of momenta near this minimum we have for $\Delta\ll ng$:
\begin{equation}   \label{g1}
g_1(r)=n_0\left[1+\frac{mg}{\pi\hbar^2}\ln\left(\frac{2ng}{\Delta}\right)J_0(2r/\xi)\right].
\end{equation}
The correction to the chemical potential due to quantum fluctuations is given by:
\begin{equation}    \label{deltamu}
\frac{\delta\mu}{\mu}\simeq \frac{2mg}{\pi\hbar^2}\ln\left(\frac{2ng}{\Delta}\right);\,\,\,\,\,\,\Delta\ll ng.
\end{equation}                                       
\begin{figure}[htb]
  \centering
  \includegraphics[scale=0.6, angle=0]{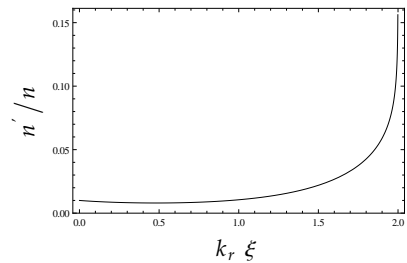}
  \caption{Non-condensed fraction as a function of $k_r\xi$ for $mg/4\pi\hbar^2=0.01$ ($\xi/r_*=100/k_r\xi$). A similar increase of the non-condensed fraction with decreasing the roton energy $\Delta$ has been found in numerical calculations of Ref. \cite{uwe}.}
  \label{mcp}
\end{figure}
      
However, the situation changes in the calculation of the compressibility. At $T=0$ the inverse compressibility is equal to $n^2\partial\mu/\partial n$. Then, using Eqs.(\ref{deltamu}) and (\ref{Delta}), for the roton minimum close to zero we obtain at $\Delta\ll ng$:
\begin{equation}    \label{comp}
\frac{\partial\mu}{\partial n}=g\left[1+\frac{2mg}{\pi\hbar^2}\ln\left(\frac{2ng}{\Delta}\right)+\frac{mg}{\pi\hbar^2}\left(\frac{2ng}{\Delta}\right)^2\right],
\end{equation}
where $g$ is the mean field contribution, and the second and third terms originate from quantum fluctuations. Small deviations of the compressibility from the mean field result require the inequality
\begin{equation}      \label{Bogcrit0}
\frac{mg}{\pi\hbar^2}\left(\frac{2ng}{\Delta}\right)^2\ll 1.
\end{equation}
We thus conclude that at $T=0$ the validity of the Bogoliubov approach is guaranteed by the presence of the small parameter (\ref{Bogcrit0}). For the dysprosium example given after Eq.(\ref{Delta}) we have $ng$ about 5 nK, and the criterion (\ref{Bogcrit0}) is satisfied for the roton energy above $2$ nK.

%\section{Fluctuations at finite temperatures}
 
In 2D at finite temperatures, long-wave fluctuations of the phase destroy the condensate \cite{merm,hoh,pop}. There is the so-called quasicondensate, or condensate with fluctuating phase. In this state fluctuations of the density are suppressed but the phase still fluctuates. The transition from a non-condensed state to quasiBEC is of the Kosterlitz-Thouless type and it occurs through the formation of bound vortex-antivortex pairs \cite{KT}. Somewhat below the Kosterlitz-Thouless transition temperature the vortices are no longer important, and in the weakly interacting regime that we consider the phase coherence length $l_{\phi}$ is exponentially large. Thermodynamic properties, excitations, and correlation properties on a distance scale smaller than $l_{\phi}$ are the same as in the case of a true BEC. Moreover, for realistic parameters of quantum gases, $l_{\phi}$ exceeds the size of the system \cite{GPS}, so that one can employ the ordinary BEC theory. 

Irrespective of the relation between $l_{\phi}$ and the size of the system, one may act in terms of the density and phase variables (hydrodynamic approach). We now show that the rotonization of the spectrum can strongly increase thermal fluctuations of the density and destroy the Bose-condensed state even at very low $T$. Using equation (\ref{deltan1}) we calculate the density-density correlation function. Assuming that the roton energy $\Delta$ is very small (at least $\Delta\ll T$), the main contribution to the integral in Eq.(\ref{deltan1}) comes from momenta near $k_r$, and we obtain:
\begin{equation}   \label{deltanT}
\frac{\langle \delta\hat n({\bf r})\delta\hat (0)\rangle}{n^2}=\frac{4mg}{\hbar^2}\frac{T}{\Delta}J_0(2r/\xi),
\end{equation}
where it is also assumed that $k_{\Delta}r\ll 1$. Comparing this result with Eq.(\ref{deltanfin}) we see that instead of the logarithmic factor we have $2\pi T/\Delta\gg 1$. 

The same factor appears in the correction to the chemical potential due to thermal fluctuations:
\begin{equation}    \label{deltamuT}
\frac{\delta\mu}{\mu}=\sum_{\bf k}(u_k-v_k)^2N_k\simeq\frac{2mg}{\hbar^2}\frac{T}{\Delta};\,\,\,\,\Delta\ll T.
\end{equation}

We now calculate the density of the normal component in the presence of the roton. In 2D the expression for this quantity reads (c.f. \cite{LL9}):
$$n_T=-\int\frac{\hbar^2k^2}{2m}\frac{\partial N_k}{\partial\varepsilon_k}\frac{d^2k}{(2\pi)^2}.$$
If the roton minimum is close to zero and $\Delta\ll T$, then the momenta near the roton minimum are the most important, and the integration over $dk$ yields:
\begin{equation}   \label{nT} 
\frac{n_T}{n}=\frac{2mg}{\hbar^2}\frac{T}{\Delta}. 
\end{equation}
The employed approach requires the condition $n_T\ll n$ because we used the spectrum of excitations obtained by the Bogoliubov method. Again, at temperatures $T\gtrsim \Delta$ we should have the inequality
$(2mg/\hbar^2)T/\Delta\ll 1$. 

A different small parameter appears in the calculation of the compressibility. The inverse isothermal compressibility is proportional to $(\partial P/\partial n)_T$, where the pressure is $P=-(\partial F/\partial S)_T$, with the free energy given by $F=E_0+T\sum_{\bf k}\ln[1-\exp(-\varepsilon_k/T)]$. For the roton minimum close to zero and $\{T,ng\}\gg\Delta$, we obtain:
\begin{equation}    \label{comprT}
\left(\frac{\partial P}{\partial n}\right)_T=ng\left[1-\frac{mg}{\hbar^2}\left(\frac{2ng}{\Delta}\right)^2\frac{T}{\Delta}+...\right],
\end{equation}
where we omitted less important finite-temperature contributions and the zero temperature contribution proportional to the small parameter (\ref{Bogcrit0}). Eq.(\ref{comprT}) shows that at $T\gg\Delta$ the Bogoliubov approach requires the inequality
\begin{equation}    \label{BogcritT}
\frac{mg}{\hbar^2}\left(\frac{2ng}{\Delta}\right)^2\frac{T}{\Delta}\ll 1,
\end{equation}
whereas for $T\lesssim \Delta$ it is sufficient to have criterion (\ref{Bogcrit0}). 

For certain quantities the Bogoliubov approach may give good results at $T\gg \Delta$ if $(mg/\hbar^2)T/\Delta\ll 1$, and at $T\lesssim\Delta$ in the presence of the ordinary small parameter $mg/2\pi\hbar^2$ amplified by a logarithmic factor $\ln(2ng/\Delta)$ for $ng\gg\Delta$. However, the validity of this approach is guaranteed only if the inequalities (\ref{BogcritT}) and (\ref{Bogcrit0}) are satisfied. For $T\gg\Delta$ the compressibility following from 
Eq.(\ref{comprT}) and the normal fraction given by equation (\ref{nT}) can become significant, by far exceeding similar quantities in an ordinary 2D Bose gas with short-range interactions at the same temperature, coupling constant $g$, and density. 
%\section{Conclusions}

In conclusion, we have shown that the roton-maxon structure of the excitation spectrum, which can be achieved in dipolar Bose-condensed gases in the 2D geometry, strongly enhances fluctuations of the density. We obtained the validity criterion of the Bogoliubov approach and found that at finite temperatures in the dilute regime where $nr_*^2\ll 1$, thermal fluctuations may significantly increase the compressibility and reduce the superfluid fraction even at temperatures well below the Kosterlitz-Thouless transition temperature.  

%\section*{Acknowledgements}

We acknowledge support from CNRS, from the University of Chlef, and from the Dutch Foundation FOM. We are grateful to L.P. Pitaevskii and D.S. Petrov for stimulating discussions.


\begin{thebibliography}{28}
\bibitem{Baranov} See for review: M. A. Baranov, Physics Reports {\bf 464}, 71 (2008).
\bibitem{Pfau} See for review: T. Lahaye et al., Rep. Prog. Phys. {\bf 72}, 126401 (2009).
\bibitem{Carr} See for review: L.D. Carr, D. DeMille, R.V. Krems, and J. Ye, New Journal of Physics {\bf 11}, 055049 (2009).
\bibitem{Pupillo2012} See for review: M.A. Baranov, M. Delmonte, G. Pupillo, and P. Zoller, Chemical Reviews, {\bf 112}, 5012 (2012).
%\bibitem{Lahaye} T. Lahaye et al., Nature {\bf 448}, 672 (2007).
\bibitem{Santos} L. Santos, G. V. Shlyapnikov, P. Zoller, M. Lewenstein, Phys. Rev. Lett. {\bf 85}, 3745 (2000).
\bibitem{Eberlein} C.Eberlein, S.Giovanazzi, D.H J O'Dell, Phys. Rev. A {\bf 71}, 033618 (2005). 
\bibitem{Santos2} L. Santos and T. Pfau, Phys. Rev. Lett. {\bf 96}, 190404 (2006).
\bibitem{pasqu} B. Pasquiou  et al., Phys. Rev. Lett. {\bf 106}, 255303 (2011).
\bibitem{ming} M. Lu et al., Phys. Rev. Lett. {\bf 107}, 190401 (2011).
\bibitem{erbium} K. Aikawa et al., Phys. Rev. Lett. {\bf 108}, 210401 (2012).
\bibitem{kk} K-K, Ni et al., Science {\bf 322}, 231 (2008).
%\bibitem{Dug} J. Deiglmayr et al., Phys. Rev. Lett. {\bf 101}, 133004 (2008).
%\bibitem{aik} K. Aikawa et al., Phys. Rev. Lett. {\bf 105}, 203001 (2010).
%\bibitem{Tak}T. Takekashi et al., Phys. Rev.A. {\bf 85}, 032506 (2012).
\bibitem{gora} L. Santos, G.V. Shlyapnikov, and M. Lewenstein, Phys.Rev. Lett. {\bf 90}, 250403 (2003).
\bibitem{gora2} G.V. Shlyapnikov and P. Pedri, Conference on correlated and Many-Body phenomena in Dipolar systems, Dresden, 2006.
\bibitem{cooper} S. Komineas, N.R. Cooper, Phys. Rev. A {\bf 75}, 023623 (2007).
\bibitem{prok} See for review: M. Boninsgni and N.V. Prokof'ev, Rev. Mod. Phys. {\bf 84}, 759 (2012).
%\bibitem{prok1} M. Boninsegni,N.V. Prokofev and B. Svistunov, Phys. Rev. E {\bf 74}, 036701 (2006).
%\bibitem{prok2} M. Boninsegni,N.V. Prokofev and B. Svistunov, Phys. Rev. Lett. {\bf 96}, 070601 (2006).
%\bibitem{Henk} N. Henkel, R. Nath and T. Pohl,Phys. Rev. Lett. {\bf 104}, 195302 (2010).
\bibitem{cint} F. Cinti, P. Jain, M. Micheli, P. Zoller and G. Pupillo, Phys. Rev. Lett. {\bf 105}, 135301 (2010).
\bibitem{bali} See for review: S. Balibar, A.D. Fefferman, A. Haziot, and X. Rojas, J. Low Temp. Phys. {\bf 168}, 221 (2012).
\bibitem{Noz} P. Nozi\`{e}res, J. Low Temp. Phys. {\bf 142}, 91 (2006); {\it ibid} {\bf 156}, 9 (2009).  
\bibitem{andr} A. F. Andreev and I. M. Lifshitz, Sov.Phys. JETP, {\bf 29}, 1107 (1969). 
\bibitem{kim} E. Kim and M. H. W T. Chan, Nature {\bf 427}, 225 (2004); Science {\bf 305}, 1941 (2004).
\bibitem{kim1} D.Y. Kim and M.H.W. Chan, Phys. Rev. Lett. {\bf 109}, 155301 (2012).
\bibitem{Pit84} L.P. Pitaevskii, JETP Letters {\bf 40}, 511 (1984).
\bibitem{Baym} G. Baym and C. Pethick, Phys. Rev. A {\bf 86}, 023602 (2012).
\bibitem{mora} C. Mora et al., Phys. Rev. B {\bf 76}, 064511 (2007).
\bibitem{bush} H. P. B\"{u}chler et al., Phys. Rev. Lett. {\bf 98}, 060404 (2007).
\bibitem{Astr} G. E. Astrakharchik et al., Phys. Rev. Lett. {\bf 98}, 060405(2007).
\bibitem{prok3} A. Filinov, N.V. Prokofev, and M. Bonitz, Phys. Rev. Lett. {\bf 105}, 070401 (2010).
\bibitem{uwe} U.R. Fischer, Phys. Rev. A {\bf 73}, 031602(R) (2006).
\bibitem{bohn1} R.M. Wilson, S. Ronen, J.L. Bohn, and H. Pu, Phys. Rev. Lett. {\bf 100}, 245302 (2008).
\bibitem{bohn2} C. Ticknor, R.M. Wilson, and J.L. Bohn, Phys. Rev. Lett. {\bf 106}, 065301 (2011). 
\bibitem{bohn3} R.M. Wilson, C. Ticknor, J.L. Bohn, and E. Timmermans, Phys. Rev. A {\bf 86}, 033606 (2012).
\bibitem{Pikgora} A. Pikovski, M. Klawunn, G.V. Shlyapnikov, L. Santos, Phys. Rev. Lett. {\bf 105}, 215302 (2010) 
\bibitem{Landlif} E. M. Lifshitz and E. M. Lifshitz, {\it Quantum Mechanics} (Heinemann, Oxford, 1999).
\bibitem{petr1} D. S. Petrov, M. Holtzmann, and G. V. Shlyapnikov,Phys. Rev. Lett. {\bf 84}, 2551 (2000).
\bibitem{petr2} D. S. Petrov and G. V. Shlyapnikov,Phys. Rev. A {\bf 64}, 012706 (2001).
\bibitem{LL9}  E.M. Lifshitz and L.P. Pitaevskii, {\it Statistical Physics, Part 2} (Pergamon Press, Oxford, 1980).
\bibitem{merm} N. D. Mermin, and H. Wagner, Phys. Rev. Lett. {\bf 22}, 1133 (1966).
\bibitem{hoh} P. C. Hohenberg, Phys. Rev. {\bf 158}, 383 (1967).
\bibitem{pop} V.N. Popov, {\it Functional Integrals in Quantum Field Theory and Statistical Physics} (D. Reidel Pub., Dordrecht, 1983).
\bibitem{KT} J.M. Kosterlitz and D.J. Thouless, J.Phys. C {\bf 6}, 1181 (1973); J.M. Kosterlitz, J. Phys. C {\bf 7}, 1046 (1974).
\bibitem{GPS} D.S. Petrov, D.M. Gangardt, and G.V. Shlyapnikov, J. Phys. IV (France) {\bf 116}, 5 (2004).


%\bibitem{Land} L. Landau, J. Phys. U.S.S.R. 11, 91 (1947).
%\bibitem{Feyn et cohen} R. P. Feynman, Phys. Rev. 94, 262 (1954).
%\bibitem{noz} P. Nozi\`{e}res, J. Low Temp. Phys. 137, 45 (2004).
%\bibitem{degenne} P.R. de Gennes, Superconductivity of Metals and Alloys (Benjamin, New York,1966).
%
%\bibitem{Kosterlitz} David R. Nelson, J. M. Kosterlitz, Phys. Rev. Lett. 39,1201 (1977).
%\bibitem{Landaulif} E. M. Lifshitz and L. P. Pitaevskii, Statistical Physics II,Pergamon Press,(1989).

%\bibitem{boudj} A. Boudjemâa and M. Benarous, Phys. Rev. A 84, 043633 (2011)
\end{thebibliography}
\end{document}